\documentclass[aps,prb,twocolumn,superscriptaddress,letterpaper,showpacs]{revtex4}
\usepackage{graphicx,bm,color,amsmath,amssymb,amsfonts,mathrsfs,txfonts,comment}

\newcommand{\al}{\alpha}
\newcommand{\be}{\beta}
\newcommand{\g}{\gamma}
\newcommand{\de}{\delta}
\newcommand{\e}{\epsilon}

\newcommand{\thi}{\theta}

\newcommand{\ka}{\kappa}
\newcommand{\la}{\lambda}

\newcommand{\p}{\pi}

\newcommand{\s}{\sigma}
\newcommand{\y}{\upsilon}
\newcommand{\f}{\phi}

\newcommand{\De}{\Delta}

\renewcommand{\y}{\psi}

\newcommand{\pd}{\partial}

\newcommand{\round}[1]{\left( #1 \right)}
\renewcommand{\square}[1]{\left[ #1 \right]}
\newcommand{\curly}[1]{\left\{#1\right\}}

\newcommand{\cvec}[2]{\round{\begin{array}{c} #1 \\ #2 \end{array}}}

\newcommand{\mat}[4]{\left(\begin{array}{cc}#1&#2\\#3&#4\end{array}\right)}

\newcommand{\rvecf}[4]{\round{\begin{array}{cccc}#1&#2&#3&#4\end{array}}}

\newcommand{\beq}{\begin{equation}}
\newcommand{\eeq}{\end{equation}}
\newcommand{\Beq}{\begin{eqnarray}}
\newcommand{\Eeq}{\end{eqnarray}}
\newcommand{\bml}{\begin{multline}}

\newcommand{\bsp}{\begin{split}}
\newcommand{\esp}{\end{split}}
\newcommand{\down}{\downarrow}
\newcommand{\up}{\uparrow}

\newcommand{\nn}{\nonumber}

\newcommand{\br}{{\boldsymbol r}}
\newcommand{\bx}{{\boldsymbol x}}

\newcommand{\bk}{{\boldsymbol k}}

\newcommand{\cH}{\mathcal{H}}
\newcommand{\hH}{\hat H}

\newcommand{\cG}{\tilde{\mathcal{G}}}
\newcommand{\cR}{\mathcal{R}}

\newcommand{\cg}{\mathcal{G}^{\rm surf}}

\newcommand{\hpmb}{\hat{\bm\s}}
\newcommand{\hpm}{\hat{\s}}

\DeclareMathOperator{\sgn}{sgn}

\newcommand{\cT}{\mathcal{T}}
\newcommand{\ve}{\varepsilon}

\begin{document}
\title{Topological superconductivity and Majorana fermions in hybrid structures \\ involving cuprate high-$T_c$ superconductors}
\author{So Takei}
\affiliation{Condensed Matter Theory Center, The Department of Physics, The University of Maryland, College Park, MD 20742-4111, USA}
\author{Benjamin M. Fregoso}
\affiliation{Condensed Matter Theory Center, The Department of Physics, The University of Maryland, College Park, MD 20742-4111, USA}
\affiliation{Joint Quantum Institute, The University of Maryland, College Park, MD 20742-4111, USA}
\author{Victor Galitski} 
\affiliation{Condensed Matter Theory Center, The Department of Physics, The University of Maryland, College Park, MD 20742-4111, USA}
\affiliation{Joint Quantum Institute, The University of Maryland, College Park, MD 20742-4111, USA}
\author{S. Das Sarma}
\affiliation{Condensed Matter Theory Center, The Department of Physics, The University of Maryland, College Park, MD 20742-4111, USA}
\date{\today}

\begin{abstract}
The possibility of inducing topological superconductivity with cuprate high-temperature superconductors (HTSC) is studied for various heterostructures.  We first consider a ballistic planar junction between a HTSC and a metallic ferromagnet. We assume that  inversion symmetry breaking at the tunnel barrier gives rise to Rashba spin-orbit coupling in the barrier and allows equal-spin triplet superconductivity to exist in the ferromagnet. Bogoliubov-de Gennes equations are obtained by explicitly modeling the barrier, and taking account of the transport anisotropy in the HTSC. By making use of the self-consistent boundary conditions and solutions for the barrier and HTSC regions, an effective equation of motion for the ferromagnet is obtained where Andreev scattering at the barrier is incorporated as a boundary condition for the ferromagnetic region. For a ferromagnet layer deposited on a (100) facet of the HTSC, triplet $p$-wave superconductivity is induced. 
For the layer deposited on a (110) facet, the induced gap does not have the $p$-wave orbital character, but has an even orbital symmetry and an odd dependence on energy. For the layer on the (001) facet, an exotic $f$-wave superconductivity is induced. We also consider the induced triplet gap in a one-dimensional half-metallic nanowire deposited on a (001) facet of a HTSC. Due to the breaking of translational symmetry in the direction perpendicular to the wire axis, the expression for the gap receives contributions from different perpendicular momentum eigenstates in the superconductor. We find that for a wire axis along the $a$-axis, these different contributions constructively interfere and give rise to a robust triplet $p$-wave gap. For a wire oriented 45$^\circ$ away from the $a$-axis the different contributions destructively interfere and the induced triplet $p$-wave gap vanishes.  For the appropriately oriented wire, the induced $p$-wave gap may give rise to Majorana fermions at the ends of the half-metallic wire. In light of the recent experimental progress in the quest for realizing Majorana fermions, we also discuss inducing superconductivity in the spin-orbit coupled nanowire using a HTSC. Based on our result, topological superconductivity in a semi-conductor nanowire may be possible given that it is oriented along the $a$-axis of the HTSC.\end{abstract}

\pacs{74.45.+c, 74.72.-h, 74.78.Na} 
\maketitle

\section{introduction}
The study of heterostructures involving a superconductor and non-superconducting materials has been
a subject of interest for many years. However, the field has recently enjoyed a renewed level of intense activity
since these systems can harbor exotic emergent particles known as
Majorana fermions.\cite{fukane,MF2,Sau2010,lutchynetal,oregetal}
These particles are sought after for their unconventional non-Abelian anyonic 
statistics and for their potential use in fault tolerant topological quantum computation.\cite{TQCrev,aliceaetalnature} 
To this date, possible realizations of Majorana fermions have been discussed in a number of systems including
$^3$He,\cite{volovik} fractional quantum Hall systems,\cite{readmoore,nayakwilczek,readgreen} organic
superconductors,\cite{senguptaetal} SrRuO$_3$,\cite{SROrev,sarmaetal} fermionic cold atoms systems,\cite{FCA1,FCA2,FCA3}
and the surface state of a topological insulator.\cite{fukane,TI1,TI2,MF2}
However, a series of theoretical works recently showed that Majorana fermions
can be realized in superconductor heterostructures with relatively conventional building blocks.\cite{Sau2010,lutchynetal,oregetal,sauetal,alicea} 
The most promising of these proposals suggests that a semiconductor nanowire fabricated on top of an $s$-wave 
superconductor, under the right conditions, should host a pair of localized Majorana excitations on the two ends of the 
wire.\cite{lutchynetal,oregetal,sauetal} The zero-bias conductance peak observed recently
can be considered as an evidence for these Majorana excitations.\cite{kouwenhoven} Very recently, transport properties
compatible with the existence of topological superconductivity has also been observed in lead nanowires.\cite{rodrigoetal}

The basis behind most of these recent Majorana proposals is to ``engineer" a single-component� $p_x+ip_y$
superconducting state in a system which does not intrinsically exhibit this property. 
In the semiconductor nanowire proposal mentioned
above,\cite{lutchynetal,oregetal,sauetal} a wire with spin-orbit coupling is placed in contact with a conventional superconductor to produce an
effective $p_x\pm ip_y$  state in the wire. Upon applying a magnetic field to the system, one of the components is
effectively removed and the desired single-component $p+ip$ topological superconductor is obtained.
Another natural possibility is to induce superconductivity in a half-metallic ferromagnet.
The system is metallic for one spin component while an insulator for the other and, as such, has a single non-degenerate
Fermi surface. If triplet $p$-wave superconductivity is induced in such a system it must harbor Majorana fermions, either in vortices
in a two-dimensional geometry,\cite{readgreen,ivanov} or at the ends in a one-dimensional wire geometry.\cite{kitaev}
Indeed, proximity-induced topological superconductivity in a half-metallic ferromagnet in contact with a conventional 
superconductor has already been studied theoretically.\cite{chungetal,duckheim11,Takei} 

Superconducting proximity effect 
in metallic and half-metallic ferromagnets in contact with conventional superconductors has received much
theoretical\cite{buzdinrev,efetovprl,SFJrev,eschrig1,eschrig2,Eschrig,haltermanPRL,haltermanPRB,bergeretPRB,braude07,
asanoPRL,asanoPRL2,fominovPRB,asanoPRB,yokoyamaPRB,linderPRB,NagaosaRev} and 
experimental\cite{SFJ1,SFJ2,HMHTSC1,keizeretal,wangetal,SFJ5,SFJ6,SFJ7,SFJ8,SFJ9}  attention over the years.
It is known that in a uniform metallic ferromagnet singlet superconducing correlations
penetrate over much shorter distances than in a non-ferromagnetic normal metal.\cite{buzdinrev} This is because 
singlet superconducting correlations compete with ferromagnetism which favors parallel alignment of electron spins.
However, early transport data reported unexpectedly long-ranged proximity effect
in Co\cite{SFJ1} and Ni\cite{SFJ2}, where the decay length for superconducting correlations 
was typical of proximity effect observed in a normal metal. 
This long-ranged proximity effect was attributed to the formation of equal-spin triplet
pairs which are immune to the pair-breaking effects of an exchange field.\cite{efetovprl,SFJrev} 
It was conjectured that local inhomogeneity in the magnetization near the S/F interface induces triplet-pairing inside 
the ferromagnet, and generates an exotic odd-frequency $s$-wave symmetry for the triplet component of the condensate
which should be robust to disorder.\cite{3he,efetovprl,SFJrev} For a half-metallic ferromagnet, 
singlet superconductivity is expected to be destroyed due to the vanishing
density of states for the minority spin component. However, supercurrent flow was remarkably observed 
between two conventional superconductors separated by CrO$_2$ (a half-metal) with a width of order $1\mu$m.
The supercurrent was attributed to odd-frequency triplet Cooper pairs which can be generated by spin-flip
scattering at junction interfaces.\cite{asanoPRL,asanoPRB}
Singlet-to-triplet conversion mechanism due to both spin-mixing and spin-flip process, 
which is inevitably present when magnetic moment near the interface is inhomogeneous, was also studied
in this context.\cite{eschrig1,eschrig2} 
Josephson coupling between a singlet and triplet superconductor in the presence of
interface spin-orbit coupling was also theoretically studied.~\cite{asanoetal03}
In the context of topological superconductivity, theoretical works have discussed another scenario for 
singlet-to-triplet conversion based on spin-orbit coupling, which can exist either in the superconductor\cite{RSOC2}
or at the S/F interface.\cite{RSOC2,RSOC3,chungetal,duckheim11,Takei} 


The reliance on the superconducting proximity effect is paramount in almost all of the current proposals for realizing topological
superconductivity.\cite{fukane,MF2,Sau2010,lutchynetal,oregetal,sauetal,alicea}
However, much of the recent interest has been in artificial creation of a simulated
$p$-wave superconductors in heterostructures where the superconductivity is proximity-induced by a conventional
$s$-wave superconductor. In contrast to all of these past works, the current work considers the creation of
topological superconductivity and Majorana fermions by using a cuprate high-$T_c$ superconductor (HTSC) 
as the agent inducing the proximity effect.
HTSCs are characterized by high transition temperatures,\cite{leggettbook} 
and may allow for more favorable experimental conditions for realizing Majorana fermions. 
In context with the preceding discussion, we focus on the physics of proximity effect between 
HTSCs and metallic (or half-metallic) ferromagnets. 
Proximity effect in HTSC-ferromagnet and/or HTSC-half-metal heterostructures have been studied theoretically with a 
combination of extended Hubbard model, Hartree-Fock theory and Bogoliubov-de Gennes equation,\cite{kuboki,zhuting,stefanakis} 
in the presence of a spin-active interface,\cite{linderPRB2,enoksen} with spin-conserved interface tunneling and spin-bandwidth asymmetry
in the ferromagnet,\cite{cuocoPRB08,annunziataPRB11} and within circuit theory.\cite{tanakaPRL03,tanakaPRB04} Enhanced 
proximity effect in the $c$-axis direction was also studied in
a diffusive ferromagnet in the presence of a domain wall near the interface.\cite{efetovprl09}
More recently, unconventional superconductor/ferromagnet junctions on the surface of a topological insulator 
have also been investigated.\cite{linder1}
On the experimental front, signatures of long-ranged Josephson coupling were observed for a 
YBa$_2$Cu$_3$O$_{7-\de}$ (YBCO)-SrRuO$_3$ (itinerant ferromagnet) 
junction.\cite{antognazzaetal} 
More recent works also showed superconducting proximity effect in thin-film heterostructures comprising of the 
half-metallic manganite La$_{2/3}$Ca$_{1/3}$MnO$_3$ (LCMO) and YBCO.\cite{HMHTSC1,HMHTSC2,HMHTSC3,kalcheimetal,HMHTSC4} 
These experiments also observed long-ranged superconducting correlations in the half-metal, implying spin-triplet pair formation in the
manganite.\cite{SFJrev,volkovetal,niuxing}

\begin{figure}[t]
\centering
\includegraphics*[scale=0.33]{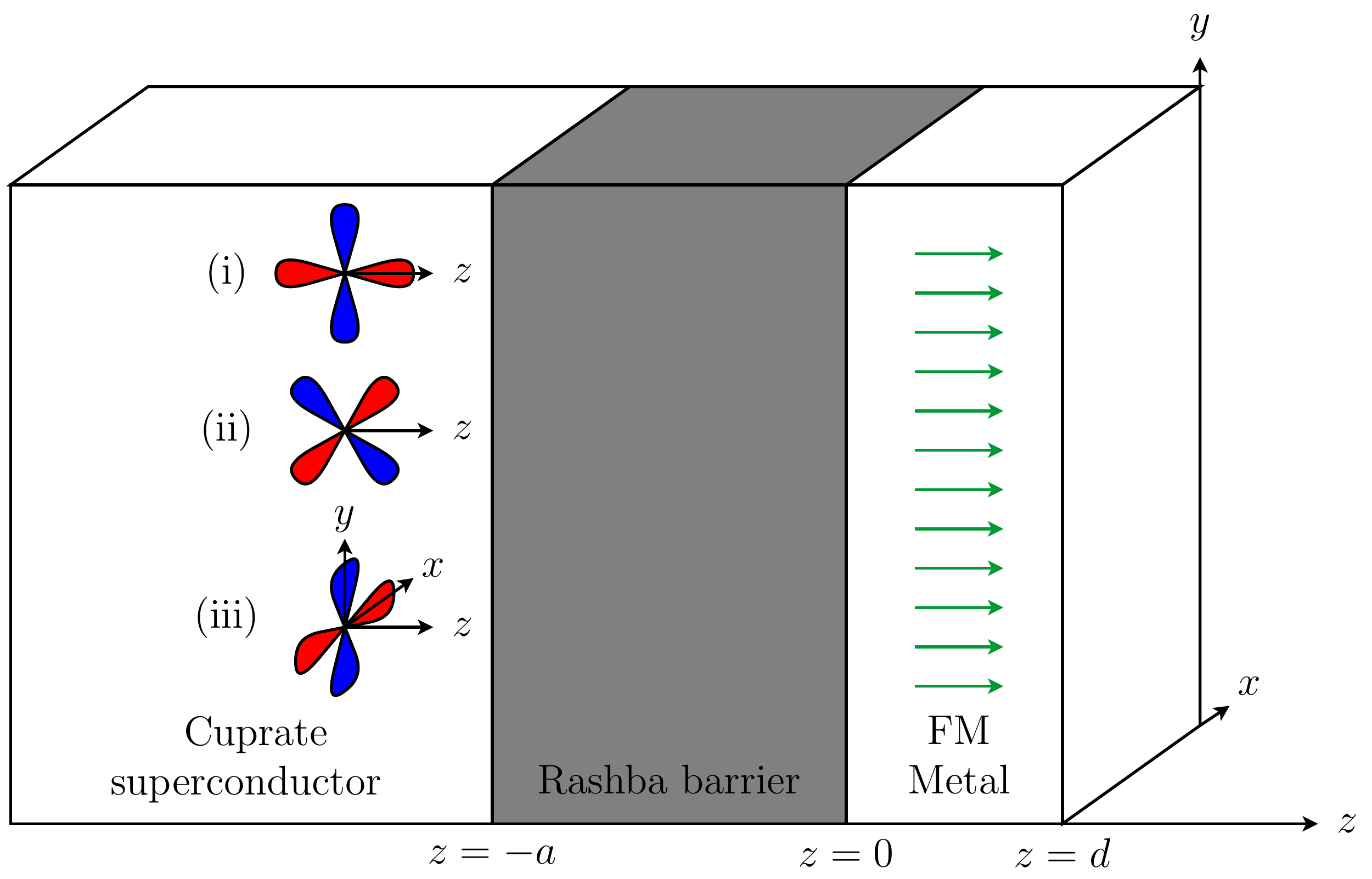}
\caption{\label{fig:sys2} (Color online) The tunnel junction considered. The ferromagnetic metal, the spin-orbit coupled barrier and 
the superconductor occupy $0<z< d$, $-a<z<0$ and $z<-a$, respectively. Various crystal orientations for the
HTSC are considered: (i) (100) parallel to the interface normal; (ii) (110) parallel to the
interface normal; and (iii) (001) parallel to the interface normal. Gap with positive (negative) sign is indicated by
red (blue). The existence of an easy axis in the $z$-direction is assumed in the ferromagnet.}
\end{figure}

Motivated by intense recent activity in the field of topological superconductivity as well as significant experimental progress 
mentioned above, we study proximity effect between HTSCs and metallic ferromagnets. 
We first consider a ballistic planar junction composed of 
the two materials separated by a potential barrier containing Rashba spin-orbit coupling.\cite{RSOC2,RSOC3,chungetal,Takei} 
The interfacial Rashba spin-orbit coupling provides the spin-flip mechanism with which equal-spin triplet pairing in 
the ferromagnet can be generated. We focus on
the proximity-induced mini-gap in the ferromagnet in the equal-spin triplet channel. In contrast to many of the
past works, we do not employ the tunneling Hamiltonian formalism to model the interface. Instead, the barrier region is 
resolved into a region of finite thickness $a$, and we microscopically model the interface with spin-orbit
coupling. We study proximity effect in both the $ab$-plane and $c$-axis directions, taking account of the hopping 
interlayer transport in the superconductor. We consider three different crystallographic orientations:
(i) $a$-axis parallel to the interface normal ((100) interface); (ii) $a$-axis 45$^\circ$ away from the interface normal 
((110) interface); and (iii) $c$-axis parallel to the interface normal ((001) interface). 
We also consider a ferromagnetic nanowire in proximity to a HTSC. 
Analytic expressions for the induced mini-gaps are provided for all the cases.
The heterostructure under consideration and the gap orientation corresponding to each of the interface orientations 
are shown in Fig.~\ref{fig:sys2}. 

An important issue when inducing topological superconductivity via proximity effect is the size of the induced mini-gap in the
normal region. This mini-gap is crucial for the protection of Majorana fermions from external perturbations.
A motivation to use a HTSC in this context is due to the robust gap it offers 
even at temperatures of order 100K. However, it is possible that the induced mini-gap amplitude is limited by the energy 
scale set by the interface spin-orbit coupling, $\De_{\rm SOC}$.
This is indeed the case for orientation (i), where the mini-gap is independent of the gap amplitude for the
proximate superconductor.
For this case, a larger gap in the superconductor therefore does not directly increase 
the size of the mini-gap. On the other hand, a proportional dependence on the gap amplitude is found for the mini-gap in
case (iii) (proximity effect in the $c$-axis direction). In this case, increasing the gap in the proximate superconductor can 
directly enhance the robustness of the gap in the normal region.
We add that even when the mini-gap does not explicitly depend on the gap of the proximate superconductor, 
it is still essential for a robust gap to be present in the superconductor to
have \textit{any} kind of gap in the normal region. In this respect, both a large $\De_{\rm SOC}$ 
and the proximity to a HTSC are a favorable combination for inducing a large mini-gap. 
Ref. \onlinecite{potterlee}
purports that certain metallic surface states can be subject to spin-orbit coupling strength of the order 
$\De_{\rm SOC}\sim 100$meV. If spin-orbit coupling of this scale can be generated in the heterostructure considered
here, having a HTSC should certainly be beneficial for generating a larger mini-gap in the
normal region. An important experimental question in realizing Majorana fermions with HTSC is the scattering 
into the nodal quasiparticle states. This issue relates to the stability of the Majorana end states, and will be
addressed in a forthcoming publication. 

The paper is organized as follows. 
The heterostructure and its theoretical modeling are introduced and discussed in Sec.~\ref{sec:model}. An effective Bogoliubov-de
Gennes equation for the ferromagnetic region is obtained in Sec.~\ref{ss:tb}. Analytical results for the proximity-induced triplet mini-gap are
discussed in Sec.~\ref{sec:gap} for various crystallographic orientations. Induced triplet mini-gap for a half-metallic
nanowire is studied in Sec.~\ref{sec:wire}. A summary and experimental implications of our work are 
presented in Sec.~\ref{summary}.

\section{The setup and Model}
\label{sec:model}
We consider a ballistic planar tunnel junction between a ferromagnetic metal ($0<z<d$) and a HTSC 
($z<-a$) as shown in Fig. \ref{fig:sys2}. The two systems are separated by a tunnel barrier ($-a<z<0$) containing 2D Rashba
spin-orbit coupling, which can arise due to breaking of inversion symmetry about the interface plane.
In the presence
of spin-orbit coupling and superconductivity it is most natural to express the equations of motion 
in terms of a four-component vector in 
Nambu-spin space, $[\Psi(\bx,t)]^T = \rvecf{\psi_{\up}(\bx,t)}{\psi_{\down}(\bx,t)}{\psi_{\up}^\dag(\bx,t)}{\psi_{\down}^{\dag}(\bx,t)}$, 
where $\bx=(\br,z)$ collects the coordinates $\br=(x,y)$ parallel to the interface plane and the transverse coordinate $z$. 
We note again that we do not employ the tunneling Hamiltonian formalism to model the interface. 
Instead, the barrier region is resolved into a region of finite thickness $a$, and we microscopically model the 
interface with spin-orbit coupling. We therefore start with equations of motion for three regions,
each of which can generally be written as 
\beq
\label{geneom}
\int d\bx'\square{\ve\hat\tau_3\de(\bx-\bx')-\cH^\rho(\bx,\bx')}\Psi^\rho(\bx',\ve)=0,
\eeq
where $\rho=\{F,B,S\}$ labels the ferromagnet, barrier and superconductor regions 
and $\ve$ is the Fourier variable for time, $\Psi^\rho(\bx,t)=\int d\ve/(2\p)\ \Psi^\rho(\bx,\ve)e^{-i\ve t/\hbar}$. 
We use $\hat\tau_i$ to denote
the vector of Pauli matrices acting on Nambu space and ``calligraphic" letters, such as $\mathcal{H}$,
to denote $4\times4$ matrices acting on Nambu-spin space. Hats will also be used in general to denote $2\times2$
matrices.

The ferromagnet is described within the standard Stoner model, assuming identical dispersion for both spins 
and with a shift in energy by the exchange interaction $h$. Assuming
that the ferromagnet has an easy axis along the $z$-axis, its Hamiltonian is given by 
$\cH^F(\bx,\bx')=\de(\bx-\bx')\cH^F(\bx)$ with
\beq
\label{HFM}
\cH^F(\bx)=\hat\tau_0\hH^F(\bx),
\eeq
where $\hH^F(\bx)=-\hbar^2\nabla_\bx^2/2m_F-\mu_F-h\hpm_3$, $m_F$ and $\mu_F$ 
are effective mass and chemical potential in the ferromagnet respectively, and $\hat\s_i$ is 
used to denote the vector of Pauli matrices acting on 
spin space. 

The barrier in between the ferromagnet and the superconductor is characterized by a large constant potential $U_0$
and Rashba spin-orbit coupling with strength $\al$. Its Hamiltonian reads $\cH^B(\bx,\bx')=\de(\bx-\bx')\cH^B(\bx)$
with
\begin{align}
\cH^B(\bx)=\mat{\hH^B(\bx)}{0}{0}{[\hH^B(\bx)]^*},
\end{align}
where $\hH^B(\bx)=-\hbar^2\nabla_\bx^2/2m_B+U_0+\alpha[\hpmb\times(-i\nabla_\bx)]_z$, and $m_B$ is the effective mass in the region.

Finally, we model the HTSC within the Bogoliubov-de Gennes mean-field theory and consider various 
orientations of its $a$-axis with respect to the interface normal. The Hamiltonian can generally be written as
\begin{align}
\cH^S(\bx,\bx')=\mat{\de(\bx-\bx')\hH^S_0(\bx)}{\hat\De(\bx-\bx')}
{\hat\De^*(\bx-\bx')}{\de(\bx-\bx')\hH^S_0(\bx)},
\end{align}
where
\beq
\label{gapmat}
\hat\De(\bx-\bx')=\mat{0}{\De(\bx-\bx')}{-\De(\bx'-\bx)}{0}.
\eeq
The dispersion in the HTSC depends on its crystallographic orientation.
We consider three different orientations where the interface is: (i) the (100) 
surface; (ii) the (110) surface; and (iii) the (001) surface (see Fig.~\ref{fig:sys2}). 
For cases (i) and (ii), we use the bulk dispersion of the form
\beq
\label{xiperp}
\hH^S_{0\perp}(\bx)=J\cos(-i\pd_x s) - \frac{\hbar^2(\pd_y^2+\pd_z^2)}{2m_S}-\mu_S,
\eeq
where $J$ is the $c$-axis (interlayer) hopping amplitude, $s$ is the distance between the copper oxide planes, and
$m_S$ and $\mu_S$ are the effective mass and chemical potential.
For case (iii) (relevant for proximity effect in the
$c$-axis direction) the relevant dispersion is
\beq
\label{xipar}
\hH^S_{0\|}(\bx)= -\frac{\hbar^2\nabla_\br^2}{2m_S} + J\cos(-i\pd_z s)-\mu_S.
\eeq
The gap function in (\ref{gapmat}) reflects the $d$-wave nature of the gap, and also depends 
on the crystal axis orientation. Its explicit expression will be given in the relevant sections below.
\begin{figure}[t]
\centering
\includegraphics*[scale=0.3]{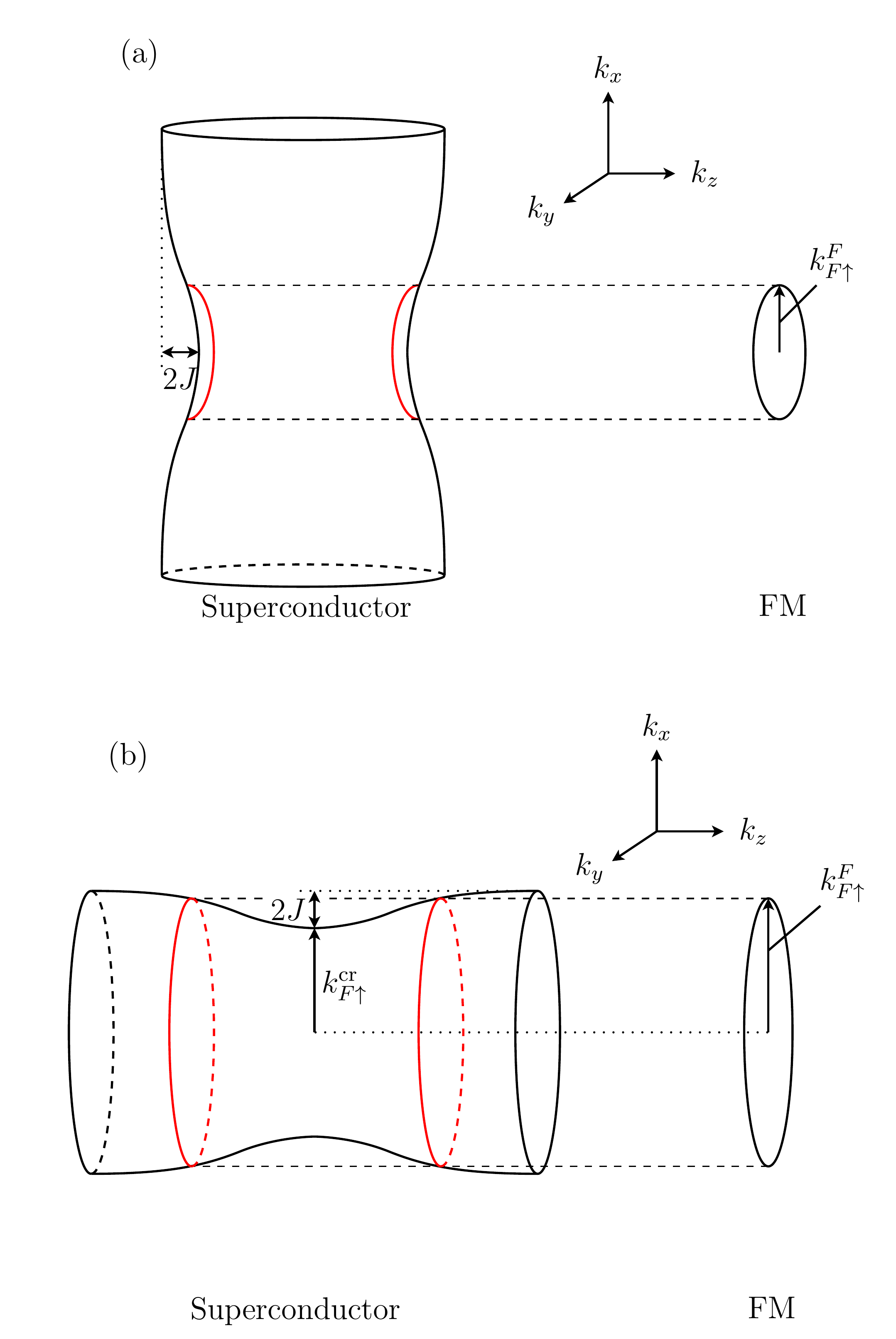}
\caption{\label{fig:FSs_perp} (Color online) (a) 3D Fermi surface corresponding to dispersion (\ref{xiperp}) on the left and 
quasi-2D Fermi surface for the majority spin component in the ferromagnet on the right. An overlap between the two
Fermi surfaces is achieved for arbitrarily small $k^F_{F\up}$. (b) 3D Fermi surface corresponding to dispersion (\ref{xipar}) on the left and 
quasi-2D Fermi surface for the majority spin component in the ferromagnet on the right. An overlap between the two
Fermi surfaces is achieved for $k^{\rm cr}_{F\up}<k^F_{F\up}<k^{\rm cr}_{F\up}+2J$.}
\end{figure}

We now discuss an assumption which will be made throughout the work (unless otherwise stated). It is an assumption regarding the 
magnitude of the Fermi wave-vector in the ferromagnet with respect to the Fermi wave-vector in the
superconductor. The outcome of this discussion can potentially influence the experimental conditions 
favorable for a robust proximity effect between these two systems.
For the clean planar tunnel junction considered in this work, momentum parallel to the interface must be conserved
during tunneling. This usually places a restriction on the region of the Fermi surface from which Cooper pairs can
tunnel.\cite{Tanaka90} In Fig.~\ref{fig:FSs_perp}(a), the 3D Fermi surface corresponding to dispersion (\ref{xiperp}) is
plotted on the left and the 2D Fermi surface for the ferromagnetic metal is plotted on the right. This corresponds to the
crystal orientations (i) and (ii) defined above (see Fig.~\ref{fig:sys2}). In Fig.~\ref{fig:FSs_perp}(b), 
the 3D Fermi surface corresponding to dispersion (\ref{xipar}) is
plotted on the left and the 2D Fermi surface for the ferromagnetic metal is plotted on the right. This corresponds to 
the crystal orientation (iii).
We are considering a quasi-2D ferromagnetic metal, which is confined in the $z$-direction,
and focus our attention on the majority spin component 
since we are interested in inducing equal-spin triplet Cooper pairs. We assume that the 
Fermi wave-vector in the ferromagnet, $k_{F\up}^F$, is controlled by externally gating the ferromagnet. Projections of the Fermi
surfaces (ring) of the ferromagnet onto the superconductor's Fermi surface are shown by the red lines.
Cooper pairs which participate in the tunneling are expected to come from these regions 
because the in-plane momenta are conserved for these processes. For orientations (i) and (ii) 
(i.e. Fig.~\ref{fig:FSs_perp}(a)), the Fermi surface with an arbitrarily small $k_{F\up}^F$ will have nonzero overlap 
with the Fermi surface of the superconductor. In this case, a robust proximity effect is expected even
when the ferromagnet is in the dilute limit (small $k_{F\up}^F$). On the other hand, for orientation (iii) 
(i.e. Fig.~\ref{fig:FSs_perp}(b)), a finite overlap with the Fermi surface of the superconductor
is achieved only for $k^{\rm cr}_{F\up}<k^F_{F\up}<k^{\rm cr}_{F\up}+2J$ within our model. Therefore, ferromagnets with Fermi wave-vectors
lying outside this region may experience a strong suppression in the proximity effect. This in principle places a restriction
on the ferromagnet's wave-vector which is favorable for strong proximity effect. In the remainder of this work, we will
assume a small $k_{F\up}^F\ll k_F^S$ for cases (i) and (ii), where $k_F^S$ is the Fermi wave-vector
in the superconductor, and $k^{\rm cr}_{F\up}<k^F_{F\up}<k^{\rm cr}_{F\up}+2J$ for case (iii).

In the following sections, Andreev physics at a HTSC-ferromagnet interface in the
presence of interface Rashba spin-orbit coupling will be studied. We note the reader that the formal procedure used to obtain
the proximity-induced mini-gaps is applicable both to a polarized as well as unpolarized normal regions. 
However, the ferromagnetism in the normal region is crucial if 
it contains multiple transverse sub-bands and one wants to realize an isolated Majorana mode,
i.e. (\# of Majorana modes) mod 2 = 1.
The usual difficulty in yielding an odd number of sub-bands at the Fermi level 
arises because the electrons have spin and consequently come in pairs.
However, an isolated Majorana mode is in principle obtainable if the normal region is a ferromagnet so that 
the total number of occupied sub-bands is odd, i.e. $(N_\uparrow + N_\down)\mbox{ mod } 2 = 1$,
and if a $p$-wave equal-spin triplet superconductivity is induced in each sub-band.


\section{Bogoliubov-de Gennes equation for the ferromagnet}
\label{ss:tb}
The first goal is to obtain the effective equation of motion for the ferromagnet. This is done by first imposing proper 
boundary conditions at each of the two interfaces, then solving the barrier and superconductor regions, and eventually 
incorporating their effects as a boundary condition for the ferromagnetic region. 
Due to translational symmetry in the $xy$-plane (discrete translational symmetry in the $x$-direction
when the copper oxide planes are stacked parallel to the $yz$ plane) we may 
introduce a Fourier variable $\bk=(k_x,k_y)$ defined via 
$\Psi^\rho_\bk(z,\ve)=\int d^2\br\ \Psi^\rho(\bx,\ve)e^{-i\bk\cdot\br}$. 
Boundary conditions are then defined in the usual way.
The first set of boundary conditions are constructed by imposing continuity of the wave-function 
at each interface,
\begin{equation}
\label{b2}
\Psi^{F}_\bk (0,\ve) =  \Psi^{B}_\bk(0,\ve),\quad 
\Psi^{B}_\bk (-a,\ve) = \Psi^{S}_\bk (-a,\ve),
\end{equation}
and vanishing of the wave-function at the hard wall,
\begin{equation}
\label{b1}
\Psi^{F}_\bk (d,\ve) = 0.
\end{equation} 
Boundary
conditions for the derivatives are obtained by integrating the equations of motion over a small 
interval with respect to $z$ and subsequently sending the interval to zero. Due to the particular
two-dimensional nature of the spin-orbit coupling, the derivative boundary conditions are simply given by
\begin{align}
\label{b3}
\frac{1}{m_F} \partial_z \Psi^{F}_\bk (0,\ve) &=  \frac{1}{m_B} \partial_z  \Psi^{B}_\bk (0,\ve),\\
\label{b4}
\frac{1}{m_B} \partial_z \Psi^{B}_\bk (-a,\ve) &=  \frac{1}{m_S} \partial_z  \Psi^{S}_\bk (-a,\ve).
\end{align}

Recall that the equation of motion for the ferromagnet is given by (\ref{geneom}) and (\ref{HFM}).
This equation of motion must be accompanied by
appropriate boundary conditions. At $z=d$, the hard wall boundary condition (\ref{b1}) applies.
The boundary condition at $z=0$ is obtained by first solving (\ref{geneom}) for the barrier and the
superconductor and by making use of the boundary conditions (\ref{b2}), (\ref{b3}) and 
(\ref{b4}).\cite{tkachov,Takei} Incorporating this boundary condition into the equation 
of motion for the ferromagnet,\cite{tkachov,Takei} we obtain
 \begin{multline}
 \label{FP}
\int dz'[\mathcal{G}^{F}_\bk(z,z',\ve)]^{-1}  \Psi_\bk^{F}(z',\ve) \\= -\de(z)\frac{\hbar^2}{m_B} 
 \left[ \cR_\bk +  \cT_\bk 
{{\hbar^2 / 2m_B} \over 1 - {\hbar^2 \over 2m_B} \cg_\bk(\ve)  \cR_\bk } \cg_\bk(\ve)
\cT_\bk \right]\Psi_\bk^{F}(z).
 \end{multline}
where the inverse Green function  $[\mathcal{G}^F_\bk(z,z',\ve)]^{-1}=\ve\hat\tau_3\de(z-z')-\cH^F_\bk(z,z')$, and $\cH^F(\bx,\bx')
=\int d^2\bk/(2\p)^2\ \cH^F_\bk(z,z')e^{i\bk\cdot(\br-\br')}$. We note that for cases (i) and (ii),
continuous translational symmetry in one of the directions parallel to the interface plane is broken down to discrete symmetry
associated with the periodic stacking of the copper oxide planes. In this case, the momentum Fourier variable corresponding
to that spatial direction must be restricted to the first Brillouin zone, $[-\p/s,\p/s]$.
The transmission and reflection matrices read
\begin{equation}
\label{BCNambu}
\cT_\bk = \begin{pmatrix}  \hat{T}_\bk & 0 \\  0&  \hat{T}_{-\bk}^* \end{pmatrix},\quad
\cR_\bk = \begin{pmatrix} \hat{R}_\bk & 0 \\  0&  \hat{R}_{-\bk}^* \end{pmatrix}.
\end{equation}
Here, the asterisk denotes complex conjugation (without transpose) and the $2\times2$ transmission
and reflection matrices are
\begin{align}
\label{T}
\hat{T}_\bk &= \begin{pmatrix} \kappa_t^+ & i \kappa_t^- e^{-i \gamma_\bk} \\  -i\kappa_t^- e^{i \gamma_\bk} & \kappa_t^+ \end{pmatrix},\\
\label{R}
\hat{R}_\bk&= \begin{pmatrix} \kappa^+ & i \kappa^- e^{-i \gamma_\bk} \\  -i \kappa^- e^{i \gamma_\bk} & \kappa^+ \end{pmatrix}.
\end{align}
We have also defined $\kappa_t^\pm = q_+ /(2 \sinh(q_+ a)) \pm q_- /(2 \sinh(q_- a))$ and
$\kappa^\pm = q_+ /(2 \tanh(q_+ a)) \pm q_-/( 2 \tanh(q_- a))$,
with $q_\pm = \square{(2 m_B/\hbar^2)\round{U_0 - \varepsilon + (\hbar^2\bk^2/2m_B) \pm \hbar\alpha |\bk| } }^{1/2}$
being the inverse penetration length of a particle with a positive/negative chirality inside the barrier, and $a$ 
is the barrier width. The angle $\gamma_\bk$ is defined via $ \bk = k \left( \cos{\gamma_\bk}, \sin{\gamma_\bk} \right)$.
Equation (\ref{FP}) is the Bogoliubov-de Gennes equation for the ferromagnet region with superconducting correlations and the
barrier spin-orbit coupling 
effects introduced via the boundary term. 
In (\ref{T}) and (\ref{R}), the coefficient $\ka^+_t$ is associated with tunneling of particles without spin flip and $\ka^-_t$ with
tunneling of particles accompanied by a spin flip. Similarly, coefficients $\ka^\pm$ are associated with the reflection of 
particles without ($+$) and with ($-$) a spin flip. As the width of the barrier grows, $\kappa_t^\pm$
decay exponentially, as expected. On the other hand, if the spin-orbit coupling in the barrier vanishes 
(i.e. $\alpha = 0$), we see that the spin-mixing terms vanish, $ \kappa^- = \kappa_t^- = 0$, and we 
recover boundary conditions for the standard proximity effect.\cite{tkachov}

In (\ref{FP}), we have also defined the ``surface" Nambu-Gor'kov 
Green function for the superconductor, which is evaluated at the interface between the barrier and the superconductor,
$\cg_\bk(\ve) = \lim_{z\to -a} \mathcal{G}^S_\bk(z,-a,\ve)$.
Here, $\mathcal{G}^S_\bk(z,z')$ is the Nambu-Gor'kov Green function matrix of the Bogoliubov-de Gennes equation 
for the superconductor valid on the half-space, $z<-a$. In the low-transparency limit, one may approximate $\mathcal{G}^S_\bk(z,z')$ 
with a Green function for an isolated superconductor. It then satisfies 
the Neumann boundary condition (zero flux condition at the right boundary), 
\beq
\label{neumannBC}
\partial_z \mathcal{G}^S_\bk(z,z')\Bigl|_{z=-a} = 0.
\eeq 
The explicit form for the surface Green function matrix $\cg_\bk(\ve)$ will depend on the crystallographic orientation. 
We will now compute the matrix next.

\subsection{Surface Green function matrix}
\label{sec:surfGF}
In this section, the surface Green function matrix $\cg_\bk(\ve)$, which appears in (\ref{FP}), is evaluated for
the three different crystallographic orientations. The Green function matrix encodes information
about the superconductor and crucially affects the Andreev physics which takes place at the junction.
The general technique
we employ to evaluate the superconductor Green function valid on the half-space $z<-a$ is the method of image 
and its generalization.\cite{Wu2002,Wu12003}

\subsubsection{Proximity effect in the (100) direction}
We begin by obtaining the surface Green function matrix $\cg_\bk(\ve)$ relevant for proximity effect in the (100) direction. 
Here, the copper oxide planes are stacked parallel to the $yz$ plane and the $a$-axis of the superconductor is oriented 
parallel to the interface normal. In this case, the superconductor possesses reflection symmetry with respect to the interface 
plane, hence one can make use of the method of image\cite{Takei,Wu2002,Wu12003} to express the Green function 
valid on the half-space in terms of the infinite bulk Green function. Similar to previous work\cite{Takei}, we have
\begin{equation}
\label{mirror}
\mathcal{G}^S_\bk(z,z',\ve) = \cG^S_\bk(z - z',\ve) + \cG^S_\bk(z + z'+2a,\ve),
\end{equation} 
where $\cG^S_{\bk}(z,\ve)$ is found by Fourier transforming the infinite bulk Nambu-Gor'kov Green function
matrix
\beq
\label{100GF}
\cG^S_\bk(z,\ve) = \int\frac{dk_z}{2\p}e^{ik_zz} 
\mat{\hat G_{\bk,k_z}(\ve)}{\hat F_{\bk,k_z}(\ve)}{\hat F_{-\bk,-k_z}(-\ve)}{\hat G_{-\bk,-k_z}(-\ve)}
\eeq 
where
\begin{align}
\hat F_{\bk,k_z}(\ve)&=\mat{0}{F_{\bk,k_z}(\ve)}{-F_{-\bk,-k_z}(-\ve)}{0},\\
\hat G_{\bk,k_z}(\ve)&=\mat{G_{\bk,k_z}(\ve)}{0}{0}{G_{\bk,k_z}(\ve)},
\end{align}
and the (time-ordered) normal and Gor'kov Green functions are defined as usual as
\begin{align}
\label{normalGF}
G_{\bk,k_z}(\ve)&=\frac{\ve+\xi^S_{\bk,k_z}}{\ve^2-[\xi^S_\perp(\bk,k_z)]^2-\Delta_{\bk,k_z}^2+i\eta},\\
\label{gorkovGF}
F_{\bk,k_z}(\ve)&=\frac{-\Delta_{\bk,k_z}}{\ve^2-[\xi^S_\perp(\bk,k_z)]^2-\Delta_{\bk,k_z}^2+i\eta},
\end{align}
where $\eta\ll1$ is a small parameter. Here, the dispersion is given by the form (\ref{xiperp}).
For this orientation the gap has the form
\beq
\label{gap100}
\Delta_{\bk,k_z}=\frac{k_z^2-k_y^2}{k_z^2+k_y^2}\Delta_0.
\eeq

We now invoke the assumption that was made in the discussion at the end of Sec.~\ref{sec:model}.
Within this assumption, most of the electrons which tunnel into the ferromagnet are normal
to the interface plane  (i.e. $|k_x|\sim|k_y|\sim k_{F,\s}^F\ll k^S_F$). Assuming also that we have strong anisotropy (i.e. $J\ll\mu_S$)
we can then approximate $\xi^S_\perp\approx \hbar^2k_z^2/2m_S-\mu_S$ and $\Delta_{\bk,k_z}\approx\Delta_0$. 
Also assuming that $|\ve|\ll\Delta_0$, we find 
\beq
\label{gorkovGint}
F_{\bk}(z,\ve)\approx\int\frac{dk_z}{2\p}\frac{\Delta_0e^{ik_zz}}{\round{{\hbar^2 k_z^2\over2m_S}-\mu_S}^2+\Delta_0^2}
\approx\frac{1}{\hbar v_F^S}.
\eeq
Here, $v_F^S$ is the Fermi velocity in the superconductor.
Here, we will drop the normal component (which can be assumed small compared to the anomalous component). 
The local Green function matrix then is given by
\beq
\label{localg100}
\cg_\bk(\ve)\approx \frac{2}{\hbar v_F^S}\hat\tau_1(i\hpm_2).
\eeq

\subsubsection{Proximity effect in the (110) direction}
We consider again the case where the copper oxide planes are stacked parallel to the $yz$ plane, but this time 
rotate the $a$-axis of the superconductor by 45$^\circ$. Here, the nodal direction points
parallel to the interface normal. The gap then has the form
\beq
\label{gapfunc110}
\Delta_{\bk,k_z}=\frac{2k_zk_y}{k_z^2+k_y^2}\Delta_0.
\eeq
Since the superconductor breaks reflection symmetry with respect to the interface plane the method of image must be
generalized.\cite{Wu2002,Wu12003} We find that the following form for the Green function provides the solution,
\begin{multline}
\label{mirror2}
\mathcal{G}^S_\bk(z,z',\ve) = \cG^S_\bk(z - z'^-,\ve) - \cG^S_\bk(z + z'^-+2a,\ve)\\
\times\square{\pd_z\cG^S_\bk(a+z'^-,\ve)}^{-1}\square{\pd_z\cG^S_\bk(-a-z'^-,\ve)},
\end{multline} 
where $z'^-=z'-0^+$.
One can verify that the Neumann boundary condition (\ref{neumannBC}) is
satisfied by (\ref{mirror2}). One can also see that the real Green function (the first term) satisfies the equation 
of motion for the Green function with a delta-function source located at $z=z'^-$ where $z'^-<-a$. The regulator $0^+>0$ makes
sure that the delta-function source is located in the appropriate half-space $z<-a$ when $z'=-a$. The image Green function (the second
term) on the
other hand satisfies the equation of motion with the delta-function source appropriately located outside the relevant half-space domain.
One can also reproduce the method of image solution (\ref{mirror}) from (\ref{mirror2}) for the case when reflection symmetry is 
present. In that case, the Nambu-Gor'kov Green function matrix obeys $\cG^S_\bk(z,\ve)=\cG^S_\bk(-z,\ve)$. Then the 
derivatives must be odd, i.e. $\pd_z\cG^S_\bk(z,\ve)=-\pd_z\cG^S_\bk(-z,\ve)$. Using this fact, (\ref{mirror2}) reduces to (\ref{mirror}) as expected.

On the surface, $z=-a$, the Green function matrix is then given by 
\begin{equation}
\label{g110}
 \cg_\bk(\ve) = \cG^S_\bk(0,\ve) 
\curly{1 - \square{\pd_z\cG^S_\bk(0^-,\ve)}^{-1}\square{\pd_z\cG^S_\bk(0^+,\ve)}}.
\end{equation}
With the same assumptions as for the (100) case, most of the electrons which tunnel through the barrier have small momenta in
the in-plane direction. One may then approximate the gap (\ref{gapfunc110}) by $\De_{\bk,k_z}\approx 2\De_0k_zk_y/(k_F^S)^2$.
We find for its normal component
\begin{align}
G_\bk(0,\ve)&\approx\int\frac{dk_z}{2\p}\frac{\round{\ve+\frac{\hbar^2k_z^2}{2m_S}-\mu_S}}
{\ve^2-\round{\frac{\hbar^2k_z^2}{2m_S}-\mu_S}^2-4\Delta_0^2\frac{k_y^2k_z^2}{(k_F^S)^4}+i\eta}\nn\\
&\approx \frac{1}{\hbar v_F^S}\frac{-i\ve}{\sqrt{\ve^2-\de_\bk^2\round{1-{\de_\bk^2\over4\mu_S^2}}+i\eta}},
\end{align}
where $\de_\bk=2\De_0k_y/k_F^S$.
Due to the oddness of the $k_z$ integral, the anomalous component $F_\bk(0,\ve)=0$.
The derivatives of the Green functions are given by
\begin{align}
\pd_zG_\bk(0^+,\ve)&\approx\int\frac{dk_z}{2\p}\frac{ik_z\round{\ve+\frac{\hbar^2k_z^2}{2m_S}-\mu_S}e^{i0^+z}}
{\ve^2-\round{\frac{\hbar^2k_z^2}{2m_S}-\mu_S}^2-4\Delta_0^2\frac{k_y^2k_z^2}{(k_F^S)^4}+i\eta}\nn\\
&\approx \frac{k_F^S}{\hbar v_F^S}\approx -\pd_zG_\bk(0^-,\ve),
\end{align}
and
\begin{align}
\pd_zF_\bk(0^+,\ve)&\approx\int\frac{dk_z}{2\p}\frac{ik_z(-\de_\bk k_z/k_F^S)e^{i0^+z}}
{\ve^2-\round{\frac{\hbar^2k_z^2}{2m_S}-\mu_S}^2-4\Delta_0^2\frac{k_y^2k_z^2}{(k_F^S)^4}+i\eta}\nn\\
&\approx -\frac{k_F^S}{\hbar v_F^S}\frac{\de_\bk}{\sqrt{\ve^2-\de_\bk^2\round{1-{\de_\bk^2\over4\mu_S^2}}+i\eta}}\nn\\
&\approx \pd_zF_\bk(0^-,\ve).
\end{align}
Then inserting the above results, we obtain
\beq
\label{GF(110)}
\cg_\bk(\ve) = \frac{-2i}{\hbar v_F^S}\frac{\ve}{\ve^2+{\de_\bk^4\over4\mu_S^2}+i\eta}
\begin{pmatrix}  \xi_\bk(\ve) &0&0&-\de_\bk \\ 0&\xi_\bk(\ve) &\de_{\bk}&0\\0&\de_{\bk}& -\xi_\bk(\ve)&0\\
-\de_{\bk} &0&0&-\xi_\bk(\ve)\end{pmatrix},
\eeq
where $\xi_\bk(\ve)=\sqrt{\ve^2-\de_\bk^2\round{1-{\de_\bk^2\over4\mu_S^2}}}$ .

\subsubsection{Proximity effect in the (001) direction}
We now consider the case where the interface normal is parallel to the $c$-axis. The gap function in this 
case depends only on the in-plane momenta and reads,
\beq
\label{dwavegapexp}
\De_\bk=\De_0\cos(2\g_\bk-2\thi_0),
\eeq
where $\thi_0$ is the angle between the $a$-axis and the $x$-axis. We will assume without loss of generality 
that $\thi_0=0$. Due to the hopping nature in the $z$-direction, 
the dispersion which must be used here is given by (\ref{xipar}), and the Fermi surface is given by a cylindrical 
shape as illustrated in Fig.~\ref{fig:FSs_perp}(b). Here, reflection symmetry with respect to the interface plane is 
retained and the trivial method of image (c.f. (\ref{mirror})) can be used. 
In the strongly anisotropic limit $J\ll\mu_S$, the normal component for the surface Green function reads 
\begin{align}
G_\bk(z=0,\ve)&\approx\int_{-\p/s}^{\p/s}\frac{dk_z}{2\p}\frac{\ve+\xi^S_\bk}
{\ve^2-\round{\xi_\bk^S}^2-\De^2_\bk+i\eta}\nn\\
&=\frac{1}{s}\frac{\ve+\xi^S_\bk}
{\ve^2-\round{\xi_\bk^S}^2-\De^2_\bk+i\eta},
\end{align}
where the in-plane dispersion $\xi^S_\bk=\hbar^2\bk^2/2m_S-\mu_S$.
For the anomalous component,
\begin{align}
F_\bk(z=0,\ve)&\approx\int_{-\p/s}^{\p/s}\frac{dk_z}{2\p}\frac{-\De_\bk}
{\ve^2-\round{\xi_\bk^S}^2-\De^2_\bk+i\eta}\nn\\
&=\frac{1}{s}\frac{-\De_\bk}
{\ve^2-\round{\xi_\bk^S}^2-\De^2_\bk+i\eta}.
\end{align}
The surface Green function matrix then reads
\begin{multline}
\label{cGF}
\cg_\bk(\ve) = \frac{2/s}{\ve^2-\round{\xi_\bk^S}^2-\De^2_\bk+i\eta}\\\times
\begin{pmatrix}  \ve+\xi^S_\bk &0&0&-\De_\bk \\ 0&\ve+\xi^S_\bk &\De_\bk&0\\
0&-\De_\bk& -\ve+\xi^S_\bk&0\\\De_\bk &0&0&-\ve+\xi^S_\bk\end{pmatrix}.
\end{multline}
The appropriate surface Green function matrix for a given orientation is inserted into (\ref{FP}). By evaluating
the right-hand side of the equation, we extract the induced triplet gap in the ferromagnet. The results are
presented and discussed in the next section.

 \section{Proximity-induced triplet gap}
 \label{sec:gap}
We now obtain analytical expressions for proximity-induced gaps in the ferromagnet, and focus
on the equal-spin triplet pairing channel.
In order to evaluate the effects of the Andreev processes at the interface, we must evaluate the
right-hand side of (\ref{FP}). The details of the superconductor are
now fully encoded in the surface Green function matrix, $\cg_\bk(\ve)$, which are evaluated for different
crystal orientations in Sec.~\ref{sec:surfGF}. 
We begin by writing the ferromagnet state vector in a separable form 
\begin{multline}
\label{product}
[\Psi_{\bk}^{F}(z,\ve)]^T\\ =  \rvecf{\f_{\bk\up}(\ve)}{\f_{\bk\down}(\ve)}{\f^\dag_{-\bk\up}(-\ve)}{\f^\dag_{-\bk\down}(-\ve)}
\y^{\rm (tr)}(z),
\end{multline}
where $ \psi^{\rm (tr)}(z)$ is the transverse envelope wave-function. We have ignored the possible spin-dependence of
this transverse wave-function since we assume $h$ to be much smaller than the Fermi energy scale in the 
ferromagnet, i.e. $h\ll\mu_F$. A discussion on obtaining 
$\y^{\rm (tr)}(z)$ is provided in Appendix \ref{app:trsol}. Here, we consider a thin enough quantum well such that most 
of the particles occupy the lowest transverse sub-band characterized by wave-function $\y_{0}^{\rm (tr)}(z)$ and 
energy $\ve_{\rm tr}^0\sim\hbar^2\p^2/2m_F d^2$.
We then introduce solution (\ref{product}) into (\ref{FP}), multiply it from the left by $\y^{\rm (tr)}_{0}(z)$, and integrate it 
over the thickness of the ferromagnet. This leads to an effective equation of motion for the 
majority spin component of the ferromagnet which generally has the form
\beq
\label{BdGp+ip}
\square{\ve-\xi^F_\bk+h}\f_{\bk\up}(\ve)-E_g(\ve,\bk)\f^\dag_{-\bk\up}(-\ve)=0.
\eeq
Here $\xi^F_\bk=\hbar^2\bk^2/2m_F-\mu_F$ is the 2D dispersion for the ferromagnet, 
$E_g(\ve,\bk)$ is the proximity-induced triplet gap, which in general can have a frequency
dependence, and we dropped a weakly induced spin-orbit coupling term and the singlet pairing term.

The task now is to obtain this proximity-induced triplet gap $E_g(\ve,\bk)$ for the various crystal orientations of the superconductor. 
Crystal axis orientation has been known to play 
an important role, for instance, in the tunneling spectra for a normal metal-HTSC
heterostructure.~\cite{TanakaRev,Bruder,Hu,Tanaka95} 
The proximity-induced triplet gap in our context also shows a strong orientation dependence.
The expression
for the gap is obtained from the right-hand side of (\ref{FP}). In computing the matrix, we keep the spin-orbit coupling
strength $\alpha$ finite only in the spin-mixing 
terms.\cite{Takei} The reflection and transmission matrices have been defined in (\ref{BCNambu}) and the surface Green functions
for various crystal orientations are evaluated in Sec.~\ref{sec:surfGF}. 
For proximity effect with the (100) surface (case (i)), the gap function has $d_{x^2-y^2}$ symmetry and 
is given by $\De^{(100)}_{\bk,k_z}=\De_0(k_z^2-k_y^2)/(k_z^2+k_y^2)$.
The induced triplet gap is given by
\begin{equation}
\label{Eg100temp}
E^{(100)}_g (\bk)= {2ie^{-i\g_\bk}fk_0 \hbar^4 \over m_B^2(1+\be^2)} \kappa_t^+ \kappa_t^- ,
\end{equation}
where $f=1/\hbar v_F^S$, $\be=m_S\ka^+/m_Bk_F^S$ and $k_0=\square{\y^{\rm (tr)}_0(0)}^2$. Here, $v_F^S$ and $k_F^S$ denote
Fermi velocity and Fermi wave-vector for the superconductor, respectively. For weak spin-orbit coupling (i.e. $\hbar\al k_{F\up}^F\ll U_0$),
$\ka^+\approx q/\tanh(qa)$ where $q\approx\sqrt{2m_BU_0/\hbar^2}$. Thus, for a high enough barrier, i.e. $qa>1$, 
$\ka^+\approx q$. We will assume here that the barrier is large but its scale is still 
smaller than the Fermi energy scale in the superconductor, i.e. $q\ll k_F^S$. In this case, $\be\ll 1$ (as long as $m_B\sim m_S$) 
and (\ref{Eg100temp}) can be approximated as
\begin{equation}
\label{Eg100}
E^{(100)}_g (\bk)\approx {2ie^{-i\g_\bk}fk_0 \hbar^4 \over m_B^2} \kappa_t^+ \kappa_t^-.
\end{equation}
We see from Eq.~(\ref{Eg100}) that the proximity-induced gap has the $p$-wave orbital character which arises due to the
presence of spin-orbit coupling in the barrier. Comparing this result with the results of Ref. \onlinecite{Takei}, (\ref{Eg100}) 
is essentially identical to the triplet gap induced by an isotropic $s$-wave superconductor. This is because for $k_{F\up}^F
\ll k_F^S$ (see discussion at the end of Sec. \ref{sec:model}), most of the electrons which tunnel 
have momentum directions that are nearly perpendicular to the interface plane. For the interface with the (100) surface 
these electrons are moving in the anti-nodal direction and are thus subject to the maximal value of the gap.
In this sense, it is reasonable that the induced triplet gap here is very similar to the proximity-induced gap with an 
$s$-wave superconductor.

We also note here that the magnitude of the induced gap is sensitive to the actual width of the quantum 
well. This sensitivity comes from its dependence on the value of the  local transverse wave-function $k_0$, which is given in (\ref{tr(0)}). 
We find that for smaller widths Andreev reflections occur multiple times and the gap is consequently enhanced.

We now move on to the case where the interface is made with the (110) surface. The gap function here has
$d_{xy}$ symmetry and is given by
$\De^{(110)}_{\bk,k_z}=2\De_0k_zk_y/(k_z^2+k_y^2)$. This situation is very different 
from the previous case. Here, most of the tunneling electrons 
have momenta in the nodal direction of the gap. We, therefore, find a suppression in the triplet gap induced on
the ferromagnetic side. The gap for this case reads (again for $\be\ll 1$)
\beq
\label{Eg110}
E^{(110)}_g(\ve,\bk)\approx-{2e^{-i\g_\bk}fk_0 \hbar^4 \over m_B^2} \frac{\ve\de_{k_y}}{\ve^2+\de^4_{k_y}/4\mu_S^2+i\eta}
\kappa_t^+ \kappa_t^-,
\eeq
where $\de_{k_y}=2\De_0k_y/k_F^S$. We see that for electrons tunneling near normal to the
interface (i.e. $k_y/k_F^S\ll 1$) the gap magnitude is strongly suppressed. Furthermore, the gap symmetry here
is not of the $p$-wave form. Apart from the $p$-wave factor $e^{-i\g_\bk}$, there is an additional factor $\de_{k_y}$.
This factor is spatially odd with respect to the $y$-direction. This is because for the (110) interface the gap function
is odd with respect to reflections in the $y$-direction. This is intimately tied to the existence of zero-bias conductance
peaks mentioned earlier. This additional factor makes the gap spatially even, but the required oddness is retained by the linear frequency factor
making this gap odd in frequency. We have also computed the \textit{singlet} gap for this interface orientation
induced in a (non-ferromagnetic) normal metal and for zero spin-orbit coupling. We find that the induced singlet gap is also 
odd in frequency. However, an \textit{even}-frequency gap results for the (100) surface interface. 
There may be a curious connection in this odd-frequency expression
for the gap and odd-frequency pairing which has been discussed extensively in the context of inhomogeneous superconductor
heterostructures.\cite{efetovprl,SFJrev,Eschrig,TanakaPRL,NagaosaRev} 

We now move on to proximity effect in the $c$-axis direction where the interface is parallel to the (001) surface. The gap
function is given by $\De_\bk=\De_0\cos2\g_\bk$, and we may assume without loss
of generality that the anti-nodal direction is along the $x$-axis (see Fig.~\ref{fig:sys2}). Here,
due to the hopping transport in the $z$-direction we do not assume $k_{F\up}^F\ll k_F^S$. As discussed at the end 
of Sec. \ref{sec:model}, it is most natural to assume that $k_{F\up}^F\sim k_F^{S,2D}$ for maximal proximity effect. For the induced triplet
gap, we find
\beq
\label{minigapc}
E^{(001)}_g(\ve,\bk)=\frac{2ie^{-i\g_\bk}sk_0\round{\hbar^2 \over m_B}^2\De_\bk\kappa_t^+ \kappa_t^- }
{\round{\hbar^2q\over m_B}^2+2s\round{\hbar^2q\over m_B}
\xi^S_\bk-s^2[\ve^2-(\xi^S_\bk)^2-\De^2_\bk]},
\eeq
where $\xi^S_\bk=\hbar^2\bk^2/2m_S-\mu_S$.
In the large barrier limit,
\beq
\label{largebarrierlimtemp}
\frac{qs\curly{\xi^S_\bk,\ \De_\bk,\ \ve}}{U_0}\ll1,
\eeq
(\ref{minigapc}) can be approximated by
\beq
\label{minigapcapp}
E^{(001)}_g(\ve,\bk)\approx 2ie^{-i\g_\bk}sk_0\De_\bk\frac{\kappa_t^+ \kappa_t^-}{q^2}.
\eeq
We can then
give an estimate for the product $q s$ in (\ref{largebarrierlimtemp}). The interlayer distance between copper oxide planes can be
estimated as $s\sim\la_1 a_0$, where $\la_1$ is a number of order $1-10$, and $a_0$ is the square lattice constant for the
copper oxide planes. But since we are assuming that the barrier energy is smaller than the Fermi energy scale 
in the superconductor, i.e. $q\sim k_F^S/\la_2$ with $\la_2>1$, the product $qs$ can be estimated as
$qs\sim\la_1 a_0/\la_2a_0=\la_1/\la_2\sim\mathcal{O}(1)$.
Therefore, the large barrier limit (\ref{largebarrierlimtemp}) can be simplified to
$\{\xi^S_\bk,\ \De_\bk,\ \ve\}\ll U_0\ll\mu_S$.
From (\ref{minigapcapp}), we see that the induced triplet gap does not have a $p$-wave form but has an $f$-wave form 
proportional to $\De_\bk e^{-i\g_\bk}$. 

We note here that the induced mini-gap (\ref{minigapcapp}) is directly proportional to the gap amplitude $\Delta_0$
in the proximate superconductor. This contrasts with the (100) case where $\De_0$ does not
explicitly enter the final expression for the mini-gap (c.f. (\ref{Eg100})). 
Therefore, for the (001) case, a larger gap in the proximate superconductor 
can directly enhance the mini-gap in the normal region.

\section{Triplet superconductivity in a half-metallic nanowire}
\label{sec:wire}
\begin{figure}[t]
\centering
\includegraphics*[scale=0.45]{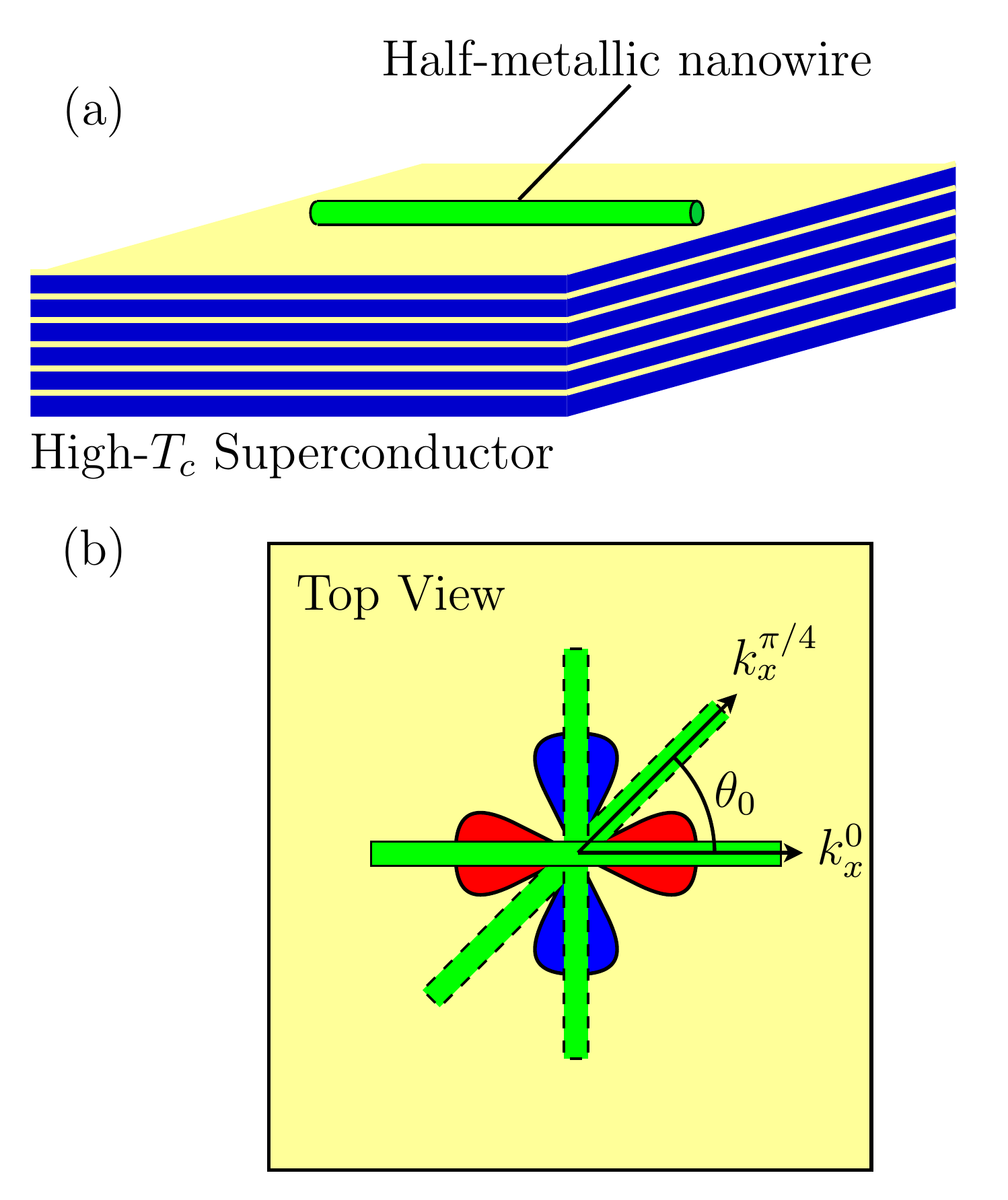}
\caption{\label{fig:wire} (Color online) (a) Half-metallic nanowire placed on the (001) surface of a HTSC. The copper
oxide planes are indicated in light yellow. (b) The nanowire placed on the superconductor with different orientations with
respect to the underlying gap symmetry. The $x$-axis is always defined along the axis of the wire.}
\end{figure}
We now consider placing a half-metallic nanowire on the (001) surface of a HTSC and
discuss the equal-spin triplet mini-gap induced in the wire. As shown in 
Fig.~\ref{fig:wire}(a) and (b), we envisage placing the nanowire in different orientations with respect to the
underlying $d$-wave gap orientation. We define the angle between the wire axis and the $a$-axis with $\thi_0$, and
consider $\thi_0=0,\p/4,\p/2$. As shown in Fig.~\ref{fig:wire}, 
we will always take the $x$-axis to be the direction along the axis of the wire. 
Due to the tunneling transport in the $c$-axis direction the physics of proximity effect
here should not depend strongly on the number of copper oxide planes considered. Therefore, we simplify the problem
by ignoring the coupling of the first copper oxide layer to the rest of the layers below, and consider the problem of
a half-metallic nanowire deposited on a 2D superconductor with a $d$-wave order parameter. In contrast to the
discussion at the end of Sec.~\ref{sec:model}, we assume that the wire is in a dilute limit where $k_{F\up}^H\ll k_F^S$, where
$k_F^S$ ($k_{F\up}^H$) is the Fermi wave-vector for the superconductor (majority spin species in the wire).

We study the equal-spin triplet gap that is induced in the half-metallic nanowire within the path-integral approach. 
Since the only focus is on obtaining an analytic expression for the triplet mini-gap the half-metallicity of the nanowire
is not strictly germane, so we relax the half-metallic condition and consider a ferromagnetic nanowire.
The Matsubara action for the 1D nanowire and the 2D superconductor can be written in the Nambu-spin basis as
\begin{align}
S_0=&\frac{1}{\hbar\be}\sum_{\e_n}\int\frac{d^2\bk}{(2\p)^2}\bar\Psi^S_\bk(\e_n)[\mathcal{G}_\bk^S(\e_n)]^{-1}\Psi^S_\bk(\e_n)\nn\\
&+\frac{1}{\hbar\be}\sum_{\e_n}\int\frac{dk_x}{2\p}\bar\Psi^W_{k_x}(\e_n)[\mathcal{G}_{k_x}^W(\e_n)]^{-1}\Psi^W_{k_x}(\e_n),
\end{align}
where the $4\times4$ inverse Green function matrix for the superconductor is given by 
\begin{align}
[\mathcal{G}_\bk^S(\e_n)]^{-1}&=-i\e_n+\hat\tau_3\xi^S_\bk+(i\hat\tau_2)(i\hat\s_2)\De_\bk,
\end{align}
and $\e_n=(2n+1)\p k_BT$ is the usual fermionic Matsubara frequency.
The Green function matrix for the ferromagnetic nanowire is given by $\mathcal{G}^W_{k_x}(\e_n)$, but we refrain from
writing down an expression for it explicitly since it will not be important in the rest of the discussion.
For $\thi_0=0$, the gap function is given by $\De_\bk^0=\De_0(k_x^2-k_y^2)/(k_x^2+k_y^2)$, while for $\thi_0=\p/4$, it reads
$\De_\bk^{\p/4}=2\De_0k_xk_y/(k_x^2+k_y^2)$. We now define the tunneling Hamiltonian. The tunnel matrix
elements will be constructed based on the results from Sec.~\ref{ss:tb}. 
However, unlike Sec.~\ref{ss:tb}, we assume that the barrier 
Rashba spin-orbit coupling resides only along the 1D boundary region between the wire and the superconductor. 
Nevertheless, the presence of spin-orbit coupling in the barrier will give rise to both spin-flip terms and a
momentum-dependence in the tunneling amplitudes. The tunneling matrix elements should reflect the same symmetry
as the 1D Rashba spin-orbit coupling present at the boundary.\cite{chungetal} We thus take them to be 
proportional to the transmission matrix given by (\ref{BCNambu}) and (\ref{T}),
\beq
\label{T1dPI}
\hat T^{1D}_{k_x}=\mat{t_{\up\up}}{it_{\up\down}\sgn(k_x)}{-it_{\up\down}\sgn(k_x)}{t_{\down\down}},
\eeq
and the tunneling Hamiltonian can be written as
\begin{align}
S_T=&\frac{1}{\hbar\be}\sum_{\e_n}\int\frac{d^2\bk}{(2\p)^2}\bar\Psi^W_{k_x}(\e_n)
\underbrace{\mat{\hat T^{1D}_{k_x}}{0}{0}{-\hat T^{1D*}_{-k_x}}}_{=:\cT^{1D}_{k_x}}\Psi^S_\bk(\e_n)+c.c..
\end{align}
The total action is then given
by $S=S_0+S_T$. At this point, we integrate out the superconductor fields and obtain an effective action
for the wire, 
\begin{align}
\label{Seffwire}
S^{\rm eff}&=\frac{1}{\hbar\be}\sum_{\e_n}\int\frac{dk_x}{2\p}\bar\Psi^W_{k_x}(\e_n)[\mathcal{G}_{k_x}^W(\e_n)]^{-1}\Psi^W_{k_x}(\e_n)\nn\\
&-\frac{1}{\hbar\be}\sum_{\e_n}\int\frac{d^2\bk}{(2\p)^2}\bar\Psi^W_{k_x}(\e_n)\cT_{k_x}^{1D}
\mathcal{G}_\bk^S(\e_n)(\cT_{k_x}^{1D})^\dag\Psi^W_{k_x}(\e_n).
\end{align}
The $k_y$-integral over the superconductor Green function matrix, $\mathcal{G}_\bk^S(\e_n)$, can now be computed.
We will now focus on the matrix component in 
the second term of (\ref{Seffwire}) which gives the equal-spin triplet mini-gap.
For $\thi_0=0$, the triplet pairing component reads
\beq
\label{egwire}
E_g^0(k_x)\approx 2it_{\up\up}t_{\up\down}f\sgn(k_x),
\eeq
where $f=1/\hbar v_F^S$. For $\thi_0=\p/4$, the integrand for the triplet pairing component becomes odd in $k_y$ 
and so we have $E_g^{\p/4}=0$. 

We see from (\ref{egwire}) that the orbital symmetry of the pairs formed in the ferromagnet is determined by the tunneling amplitudes
defined in (\ref{T1dPI}). The pairing occurs within this model because electrons with momenta $k_x$ and $-k_x$ in the wire hop 
into the superconductor, where they form a pair. Because of translational symmetry in the $x$-direction the momentum component
$k_x$ must be conserved during hopping. On the other hand, due to translational symmetry breaking in the $y$-direction a pair
of electrons with momenta $k_x$ and $-k_x$ can form a pair inside the superconductor with many $k_y$ values, i.e. Cooper pairs
with momenta $(k_x,k_y)$ and $(-k_x,-k_y)$ where $k_y$ can take on many values. The effective gap produced in the
ferromagnet is found by summing over all of these possible Cooper pairs. 
\begin{figure}[t]
\centering
\includegraphics*[scale=0.45]{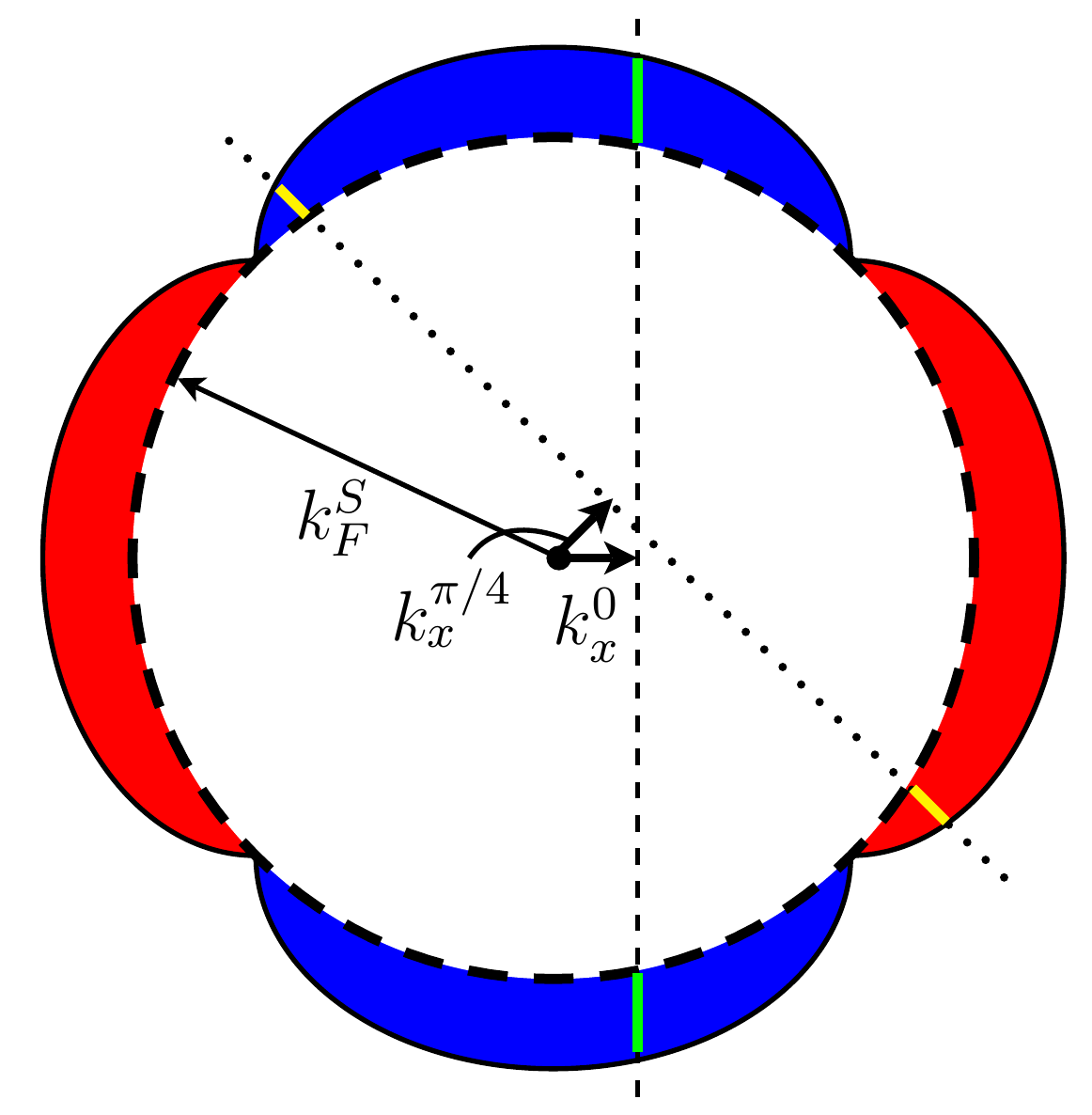}
\caption{\label{fig:SCFS} (Color online) The 2D Fermi surface of the superconductor (thick dashed line) 
and the  $d$-wave gap are schematically
plotted. Red (blue) region corresponds to the positive (negative) part of the gap. The Fermi wave-vector in the
superconductor is given by $k_F^S$. For a dilute wire (i.e. $k_{F\up}^H\ll k_F^S$), the momentum of a typical particle in the wire
is also plotted for $\thi_0=0$ and $\thi_0=\p/4$ (the short arrows). The $x$-axis is always defined to be along the axis of the
wire. Due to translational
symmetry breaking in the $y$-direction, momentum in that 
direction is not conserved during tunneling. In contrast, due to translational symmetry in the $x$-direction,
$k_x$ must be conserved during tunneling. 
The dashed (dotted) line represents regions in momentum space
which can be explored by a particle as it tunnels from the wire into the superconductor. 
For $\thi_0=0$ regions in momentum space which can contribute to the induced gap in the wire
is colored in green. For $\thi_0=\p/4$, it is colored in yellow.}
\end{figure}

In Fig.~\ref{fig:SCFS}, the Fermi surface of the superconductor and the $d$-wave gap are schematically plotted. 
Assuming that we are at low temperatures, $T\ll\De_0$, 
an electron tunneling into the superconductor from the wire has momentum $k_x\sim k_{F\up}^H$. For $\thi_0=0$,
the orientation and amplitude of $k_x$ with respect to the Fermi surface of the superconductor is also plotted
in Fig.~\ref{fig:SCFS} and labeled as $k_x^0$. For $\thi_0=\p/4$, it is labeled as $k_x^{\p/4}$. The dashed and dotted lines are used to
represent momenta a particle can explore as it tunnels into the superconductor from the half-metal. The dashed line
is for $\thi_0=0$ and the dotted line is for $\thi_0=\p/4$. For $\thi_0=0$, the summation over $k_y$
involves regions in momentum space where the gap has the same sign (green line segments in Fig.~\ref{fig:SCFS}).
In this case, the different regions give constructive interference and a robust gap is obtained.
On the other hand, for $\thi_0=\p/4$, the summation over $k_y$ involves regions in momentum space
where the gap has equal but opposite sign (yellow line segments in Fig.~\ref{fig:SCFS}). In this case, the two regions destructively
interfere and the gap vanishes. With the same token, strong proximity effect is expected again for $\thi_0=\p/2$, but the induced gap will have a sign
opposite to the $\thi_0=0$ case. 
For $\thi_0=0$ and $\thi_0=\p/2$, the induced gap in the wire has the $p$-wave orbital character (see (\ref{egwire})) rendering
the realization of Majorana fermions on the ends of the half-metallic wire possible.\cite{kitaev}

Now let us consider the case where the half-metallic wire is replaced by a non-ferromagnetic nanowire but with
spin-orbit coupling. We now remove the spin-orbit coupling in the barrier (i.e. $t_{\up\down}=0$ in (\ref{T1dPI})),
and the resulting induced gap in the wire should have an $s$-wave orbital character. 
For a wire with the $\thi_0=0$ orientation the induced $s$-wave gap has the form
\beq
E_g^{\rm singlet}(k_x)=-2t_{\up\up}t_{\down\down}f,
\eeq
while the gap vanishes again for $\thi_0=\p/4$. We note that the $\thi_0=0$ wire has the essential ingredients
for realizing  Majorana fermions on its ends. 

\section{Summary and experimental implications}
\label{summary}
In this work, we study proximity effect between a HTSC and a metallic ferromagnet.
We first consider a quasi-2D ferromagnet deposited on (100), (110), and (001) facets of the superconductor.
The induced equal-spin triplet mini-gap for each of the geometries is analytically evaluated. We assume that the
breaking of mirror symmetry at the boundary region between the superconductor and ferromagnet gives rise to
Rashba spin-orbit coupling in the region. In contrast to many of the
past works, we do not employ the tunneling Hamiltonian formalism to model the interface. Instead, the barrier region is 
resolved into a region of finite thickness, and the interface with spin-orbit coupling is microscopically modeled.
We also consider a half-metallic nanowire placed on top of a (001) facet of a HTSC. 
The induced triplet mini-gap in the wire is obtained for various wire orientations with respect to the underlying orientation of
the $d$-wave order parameter. We now highlight some relevant transport and proximity effect experiments involving high-$T_c$ cuprates,
and then comment on some experimental implications based on the findings in this work.

The in- and out-of-plane transport anisotropy is also known to affect the proximity effect involving HTSCs.
Experimentally, proximity effect was found to be weaker in the $c$-axis direction than in the $ab$-plane 
direction due to a much shorter coherence length in the former direction.\cite{gijsetal,durusoyetal,sharonietal} 
We note, however, that so-called \textit{giant proximity effect} has been reported both in the 
$ab$-plane\cite{kabasawaetal,tarutanietal,barneretal,deccaetal} and $c$-axis\cite{beasleyprl} directions. 
Proximity effect in HTSC-ferromagnet and/or HTSC-half-metal heterostructures have also been studied experimentally.
Early studies showed long-ranged Josephson coupling for a YBCO-SrRuO$_3$ (itinerant ferromagnet) 
junction,\cite{antognazzaetal} though the results may be controversial.\cite{asulinetal06}
More recent works showed superconducting proximity effect in thin-film heterostructures comprising of the half-metallic manganite La$_{2/3}$Ca$_{1/3}$MnO$_3$ (LCMO) and YBCO.\cite{HMHTSC1,HMHTSC2,HMHTSC3,kalcheimetal,HMHTSC4} 
These experiments also observed long-ranged superconducting correlations in the half-metal, implying spin-triplet pair formation in the
manganite,\cite{efetovprl,SFJrev,volkovetal,niuxing} but further verification may still be needed.\cite{HMHTSC5}
A very recent work has shown evidence for MacMillan-Rowell resonance in the same LCMO-YBCO heterostructure, further 
substantiating the evidence for long-ranged proximity effect in this system.\cite{HMHTSC4} 

We first comment on heterostructures built from a 2D metallic ferromagnet (or half-metal) 
deposited on one of the facets of a HTSC. 
Here, our results are likely to
be applicable when the Fermi wave-vector for the majority spin species in the ferromagnet is much less than 
the Fermi wave-vector of the superconductor (see discussion at the end of Sec.~\ref{sec:model}). 
We find that a robust spin-triplet $p$-wave superconductivity 
is induced in a 2D half-metal if the layer is deposited on the (100) facet of the HTSC 
(c.f. (\ref{Eg100})). On the other hand, depositing the same layer on the (110) facet is expected to be less favorable 
from the point of view of obtaining a robust gap (c.f. (\ref{Eg110})). This is because most of the electrons which tunnel 
from the superconductor into the ferromagnet have momenta in the nodal direction of the gap. As such, these 
electrons are not subject to the gap in the superconductor and a weaker gap is therefore induced in the ferromagnet.
Furthermore, in this orientation, the induced gap does not have a $p$-wave orbital symmetry, but has an even
orbital symmetry. The required overall oddness is maintained via an odd-dependence on energy. For a half-metal
layer deposited on a (001) facet, the situation differs from the other two cases. In this case,  a robust proximity effect
is expected when the Fermi wave-vector in the half-metal is approximately equal to the (2D) Fermi wave-vector for electrons in a 
copper oxide plane (see discussion at the end of Sec.~\ref{sec:model}). The induced gap here has an $f$-wave
orbital symmetry. Therefore, from the point of view of obtaining a robust $p$-wave mini-gap,
depositing the ferromagnetic layer on a (100) facet of the superconductor is the most favorable option.

The expressions for the mini-gaps obtained in Sec.~\ref{sec:gap} have been assumed to be uniform in the
normal region due to strong transverse confinement.
For a clean, thin normal region, a robust $s$-wave singlet mini-gap co-exists with the triplet gap for the (100)
interface, and is 
given by $E_g^{\rm sing}=(2fk_0 \hbar^4 / m_B^2) (\kappa_t^+)^2$ (we obtain no $S_z=0$ triplet mini-gap
within our model). These mini-gap expressions are essentially valid also for an unpolarized metal since the existence 
of the triplet gap relies only on the presence of the Rashba spin-orbit coupling at the interface. However, as mentioned
at the end of Sec.~\ref{sec:model}, the ferromagnetism in the normal region becomes crucial for realizing an isolated 
Majorana zero mode at the boundary of the region.

One may also consider depositing a half-metallic nanowire on top of a HTSC. From the
experimental point of view, perhaps the most natural geometry is to place the wire on top of a (001) facet of the 
superconductor. This is indeed the orientation we consider here, and we show that inducing an equal-spin triplet $p$-wave gap in 
the wire is indeed possible. However, one must note that the $p$-wave gap will not be induced in the wire
when it is oriented along the nodal direction of the underlying gap, i.e. the wire axis is 45$^\circ$ away from
the $a$-axis (see Fig.~\ref{fig:wire} and discussion in Sec.~\ref{sec:wire}). The absence of the gap is due to 
destructive interference among different hopping processes which contribute to the proximity effect. For the wire 
orientation along the $a$- or the $b$-axis, the different hopping processes constructively interfere and a 
robust gap is expected.

We also add that this setup used for the half-metallic nanowire can also be applied to the case where a non-ferromagnetic spin-orbit coupled nanowire is placed on top of a (001) facet of the HTSC. This setup is more akin to the recent semi-conductor nanowire proposal for realizing Majorana fermions.\cite{kouwenhoven,lutchynetal,oregetal} We may apply the above nanowire study to the case where the Rashba spin-orbit coupling is introduced in the nanowire instead of the barrier region. In this case, for a wire oriented in the $a$-axis direction, an $s$-wave gap should be induced in the wire (see discussion in Sec.~\ref{sec:wire}). The gap, however, should vanish (or be strongly suppressed) when the wire is oriented 45$^\circ$ away from the $a$-axis. The system considered here can serve as a starting point for the recent Majorana proposal presented in Refs. \onlinecite{lutchynetal} and \onlinecite{oregetal}. An important experimental question in realizing Majorana fermions with HTSC is the scattering 
into the nodal quasiparticle states. This issue relates to the stability of the Majorana end states, and will be
addressed in a forthcoming publication. 
 
While our discussion for the nanowire geometry was restricted to a strictly one-dimensional wire, recent theoretical works
have shown that this one-dimensionality condition is not required for realizing Majorana fermions.\cite{Q1D1,Q1D2,Q1D3,Q1D4}
In Refs. \onlinecite{Q1D1} and \onlinecite{Q1D2} it was shown that Majorana end states can still be realized if the wire width
does not greatly exceed the superconducting coherence length and an odd number of transverse sub-bands are occupied. Furthermore,
Refs. \onlinecite{Q1D3} and \onlinecite{Q1D4} studied how the stability of the topological superconducting state can
be enhanced in multi-band wires.

Our findings suggest that for suitably designed hybrid structures and with good contact established at the interfaces 
Majorana fermions can be realized at the ends of either a ferromagnetic 
or a spin-orbit coupled semi-conductor nanowire in contact with a HTSC.
If a vortex can be created in a 2D geometry, a Majorana fermion will reside at the vortex core while the other will reside 
at the outer edge.  The manipulation of this state is less well-studied but still interesting.

\subsection*{Acknowledgment}
This work was supported by DOE DESC0001911 (S. T.), JQI-PFC (B. M. F.), NSF-CAREER award 0847224 (V. G.) and DARPA-QuEST and Microsoft Q (S. D. S.).
One of the authors (S. T.) gratefully acknowledges helpful discussions with A. Schnyder during the early stages of this work.

\appendix

\section{Obtaining the transverse wave-function}
\label{app:trsol}
In this appendix, we evaluate the transverse wave-function for the particles residing in the quasi-2D ferromagnet. 
This quantity is used in (\ref{product}) and appears in all the expressions for the induced mini-gap.
We will assume
here that the exchange energy scale is smaller than the Fermi energy, $h\ll \mu_F$. In principle,
the transverse wave-functions for the majority and minority species are different. For instance,
in the limit where $h\gg\mu_F$, we expect that the transverse wave-function for the minority
component will be an evanescent wave which decays with the scale $\ell_h\approx[2m_Fh/\hbar^2]^{-1/2}$.
For $h\ll\mu_F$, however, the dependence can be ignored.

The transverse wave-function can be approximated as a solution to the
free 1D Sch{\"o}dinger equation, 
\begin{equation}
\label{tr}
-\frac{\hbar^2\partial^2_z}{2m_F} \y^{\rm (tr)}(z)=\varepsilon_{\rm tr}\psi^{\rm (tr)}(z),
\end{equation}
with the boundary conditions,
\beq
\label{trBC}
{ m_B \over m_F} \partial_z \psi^{\rm (tr)}(0)  =  \kappa^+ \psi^{\rm (tr)}(0),\quad  \psi^{\rm (tr)}(d)  =0.
\eeq 
The boundary value of the transverse wave function can be obtained from (\ref{tr}) and reads
 \begin{equation}
 \label{tr(0)}
k_0 =\square{\y^{\rm (tr)}_0(0)}^2= {2 \over d} {1 \over 1 + \xi^2 + \xi/(k_{\rm tr} d)},
 \end{equation}
where $\xi = m_F q/(m_B k_{\rm tr})$ and $k_{\rm tr}$ is a solution to the eigenvalue problem, 
$\tan{(k_{\rm tr} d)}/(k_{\rm tr} d) = - m_B / (m_F qd)$, which determines the spectrum, 
$\varepsilon_{\rm tr} = \hbar^2 k_{\rm tr}^2/(2m_F)$.  For the lowest sub-band solution, $\y^{\rm (tr)}_0(z)$, the
energy is given by $\varepsilon_{\rm tr}^0 = \hbar^2 (k_{\rm tr}^0)^2/(2m_F)$, where $k_{\rm tr}^0$ is the wave-vector
corresponding to the lowest energy sub-band.

We note that $k_0$ is proportional to $1/d$ where $d$ is the width of the transverse confinement. As the
width decreases $k_0$ increases, signifying the fact that the particles reflect back to the interface more
frequently and, thus, has a higher probability of residing there. This amplifies the Andreev physics at the boundary
and leads to the enhancement of the induced mini-gap. This was briefly discussed in Sec.~\ref{sec:gap}.

\bibliography{dwave_proximity}

\end{document}